# Room temperature magnetic phase transition in an electrically-tuned van der Waals ferromagnet

Cheng Tan[1,4#], Ji-Hai Liao[2#], Guolin Zheng[3*], Meri Algarni[4,5], Jia-Yi Lin[2], Xiang Ma[6], Edwin L. H. Mayes[7], Matthew R. Field[7], Sultan Albarakati[4,8], Majid Panahandeh-Fard[4], Saleh Alzahrani[4], Guopeng Wang[9], Yuanjun Yang[1], Dimitrie Culcer[10], James Partridge[4], Mingliang Tian[3,9], Bin Xiang[6*], Yu-Jun Zhao[2*] and Lan Wang[1,4*]

[1]Lab of Low Dimensional Magnetism and Spintronic Devices, School of Physics, Hefei University of Technology, Hefei, Anhui 230009, China.

[2]Department of Physics, South China University of Technology, Guangzhou 510640, China.

[3]Anhui Province Key Laboratory of Condensed Matter Physics at Extreme Conditions, High Magnetic Field Laboratory, Chinese Academy of Sciences (CAS), Hefei, Anhui 230031, China.

[4]ARC Centre of Excellence in Future Low-Energy Electronics Technologies (FLEET), School of Science, RMIT University, Melbourne, VIC 3001, Australia.

[5] Physics Department, Faculty of Science, Al-Baha University, Alaqiq 65779, Saudi Arabia.

[6]Hefei National Laboratory for Physical Sciences at the Microscale, Department of Materials Science & Engineering, CAS Key Lab of Materials for Energy Conversion, University of Science and Technology of China, Hefei, Anhui 230026, China.

[7]RMIT Microscopy & Microanalysis Facility, RMIT University, Melbourne, VIC 3000, Australia.

[8]Physics Department, Faculty of Science and Arts, University of Jeddah, P.O. Box 80200, 21589Khulais, Saudi Arabia.

[9]Department of Physics, School of Physics and Materials Science, Anhui University, Hefei, Anhui 230601, China.

[10]School of Physics and ARC Centre of Excellence in Future Low-Energy Electronics Technologies, UNSW Node, University of New South Wales, Sydney, New South Wales 2052, Australia.

[#] These authors equally contributed to the paper.

[*] Corresponding authors. Correspondence and requests for materials should be addressed to G. Z. (email: glzheng@hmfl.ac.cn), B. X. (email: binxiang@ustc.edu.cn), Y.-J. Z. (email: zhaoyj@scut.edu.cn) and L. W. (email: wanglan@hfut.edu.cn).

## Abstract

Finding tunable van der Waals (vdW) ferromagnets that operate at above room temperature is an important research focus in physics and materials science. Most vdW magnets are only intrinsically magnetic far below room temperature and magnetism with square-shaped hysteresis at room-temperature has yet to be observed. Here, we report magnetism in a quasi-2D magnet $Cr_{1.2}Te_2$ observed at room temperature (290 K). This magnetism was tuned via a protonic gate with an electron doping concentration up to $3.8 \times 10^{21}$ cm$^{-3}$. We observed non-monotonic evolutions in both coercivity and anomalous Hall resistivity. Under increased electron doping, the coercivities and anomalous Hall effects (AHEs) vanished, indicating a doping-induced magnetic phase transition. This occurred up to room temperature. DFT calculations showed the formation of an antiferromagnetic (AFM) phase caused by the intercalation of protons which induced significant electron doping in the $Cr_{1.2}Te_2$. The tunability of the magnetic properties and phase in room temperature magnetic vdW $Cr_{1.2}Te_2$ is a significant step towards practical spintronic devices.

Electrical manipulation of magnetic properties is essential for the development of low-energy spintronic devices.[1-5] In recent years, the emergence of van der Waals (vdW) magnets has increased the synergy between spintronics and two-dimensional (2D) materials.[6-11] For the next generation of low-energy vdW spintronic devices, electrically tunable magnetic properties will be vital. To date, various electrically tuned magnetic properties, such as magnetization [12], coercivity [13], magnetic anisotropy [14], magnetic phase transition (MPT) [15,16], and exchange bias [17,18], have been observed in vdW magnets and heterostructures. However, these phenomena have only been observed at low temperatures. Recently, chromium chalcogenides ($Cr_aX_b$, X = S, Se and Te), a class of layered or quasi-layered structure magnets, have exhibited interesting and potentially exploitable magnetic characteristics at higher temperature [19-28]. Among them, $Cr_aTe_b$ materials have exhibited high Curie temperatures and more importantly, remanence and perpendicular magnetic anisotropy that appear to remain significant (i.e. near square-shaped hysteresis loop exhibited) at room temperature [29]. These materials thus hold promise for future applications and may be suitable platforms for room temperature studies of electric gate-controlled magnetism. However, $Cr_aTe_b$ materials are metallic and thus difficult to electrically tun using standard dielectric layers. Ionic gating techniques have recently proved successful in elevating the Curie temperature of $Fe_3GeTe_2$ to room temperature while the low remanence at room temperature limits its broader application in spintronics [30]. Notably, a solid protonic gating technique [16-18,31] has proved effective in altering the magnetic characteristics of vdW metals.

Here, for the first time, we report an electrically tuned phase transition from ferromagnetism (FM) to antiferromagnetism (AFM) in a vdW magnet with near square-shaped loops at room temperature (290 K), using proton-gated $Cr_{1.2}Te_2$ nanoflakes. In this work, the magnetism of a pristine $Cr_{1.2}Te_2$ nanoflake was firstly investigated under magnetic fields applied perpendicular to its surface. These experiments revealed square-shaped anomalous Hall effect (AHE) loops which persisted at temperatures exceeding 50 K and indicated the existence of a perpendicular magnetic anisotropy. Solid protonic gates were then added and these were used to modulate the magnetism of $Cr_{1.2}Te_2$ nanoflakes at 200 K. The results of these experiments demonstrated that electron doping concentrations up to $3.8 \times 10^{21}$ cm$^{-3}$ could be achieved with proton intercalation. The AHE resistivity ($\rho_{AHE}$) and coercivity were effectively tuned with gate voltage. Finally, the magnetism of a $Cr_{1.2}Te_2$ nanoflake was electrically tuned at 290 K, resulting in coercivity and $\rho_{AHE}$ exhibiting acute evolution that resembled the characteristics measured at 200 K. The AHE loops of all $Cr_{1.2}Te_2$ nanoflake devices vanished under high gate-voltages. This behaviour is attributed to MPT. Supporting DFT calculations indicate that a transition from FM to AFM occurs due to proton-intercalation-induced electron doping within the $Cr_{1.2}Te_2$. The calculated anomalous Hall conductivity (AHC) varied significantly near the Fermi level, explaining the dramatic evolution of gate-voltage dependent AHE. Combined the experimental work and calculations shed light on the underlying mechanisms and prospects for electrically tuned vdW spintronic devices.

As a newly synthesized room-temperature vdW magnet, $Cr_{1.2}Te_2$ is noteworthy due to its high remanence near 290 K and the perpendicular magnetic anisotropy up to 300 K [29]. In this study, several $Cr_{1.2}Te_2$ devices were fabricated and subjected to measurements. These are labelled as device #1, device #2, *etc.* Figure 1(a) shows the lattice structure of $Cr_{1.2}Te_2$ which has a space group of $P\bar{3}m1$ and exhibits the same intralayer structure (CrTe layer) as $CrTe_2$, while its covalently bonded octahedral vacancies in the vdW gaps are fractionally occupied by the Cr atoms [Cr-2 in Figure 1(a)]. As the vacancy occupancy rate is ~20% in $Cr_{1.2}Te_2$, the average covalent bond between the adjacent CrTe layers is weaker than the intralayer bonds. This facilitates the mechanical exfoliation of the $Cr_{1.2}Te_2$ nanoflakes. Comprehensive characterization revealed that the Cr/Te ratio is homogeneous throughout the crystals/nanoflakes (see Fig. S1). Figure 1(b) shows an optical image of a $Cr_{1.2}Te_2$ nano-device

(device #1). The atomic force microscopy [Figs. 1(c) and (d)] indicates a thickness of around 47 nm. Figure 1(e) shows the AHE curves from 2 K to 200 K when the magnetic field is perpendicular to the surface of device #1. The $\rho_{AHE}$ values (defined as $2 \times |\rho_{xy}(B = 0.1\ T)|$) of each curve are indicated. At 2 K and 20 K, the hysteresis loops deviate from square-shaped. From the previous study [29], the as-grown $Cr_{1.2}Te_2$ crystal exhibits in-plane magnetic anisotropy at lower temperatures, while its magnetization easy axis can be rotated from the in-plane to out-of-plane direction when the temperature exceeds 140 K. Here, akin to previous results, near-square shaped hysteresis loops are indicative of perpendicular magnetic anisotropy when the temperature is over 50K. Figure 1(f) shows the temperature dependent AHE curves recorded above 250 K. These curves became absent at around 320 K, confirming the high Curie temperature of $Cr_{1.2}Te_2$. The temperature dependent coercivity and remanence curves from 2 K to 300 K are displayed in Fig. 1(g). Here, the coercivity is the value of the magnetic field at $\rho_{xy}$=0 and the remanence is the ratio of $\rho_{xy}$ (B = 0 T, decreased from 2T)/$\rho_{xy}$ (B = 0.1 T) (i.e. $R_0/R_S$ ratio). Owing to the in-plane magnetic anisotropy, near-zero coercivities and very small remanences were observed at 2 K and 20 K. When the temperature exceeded 50 K, the coercivity first increased with increasing temperature before reaching a peak of 550 Oe at around 200 K. Thereafter, the coercivity decreased with increasing temperature before vanishing when T exceeded 300 K. From 50 K to 280 K, the remanence maintained a value of around 1, implying a dominant perpendicular magnetic anisotropy. From 280 K, the remanence started to decrease and finally reached 0 at 300 K, indicating that thermal agitation was then dominant. In this study, we primarily focus on nanoflakes with a thickness of approximately 40 nm. Compared to other thicknesses (see Fig. S3), their AHE loops maintained a more square shape at higher temperatures.

As stated prior, effective control of the magnetism by a gate voltage represents a significant step in broadening the range of applications for spintronics. Here, using a protonic gate technique, we demonstrate that the ferromagnetism in $Cr_{1.2}Te_2$ nanoflakes can be electrically modulated at temperatures up to and including room temperature. As shown in Fig.1, the $\rho_{AHE}$, coercivity and $R_0/R_S$ ratio of device#1 reached a peak around 200 K, hence we firstly investigated the gate tuned AHE loops at 200 K. Figure 2(a) shows a schematic of a solid protonic gate device, in which a voltage is applied between the Pt gate electrode (under the solid protonic electrolyte) and the $Cr_{1.2}Te_2$ nanoflake to drive protons into the nanoflake. Figure 2(b) shows the evolution of the $\rho_{xy}$ loops of a 43 nm thick $Cr_{1.2}Te_2$ nanoflake (device #8) under various gate voltages ($V_g$). When the $V_g$ was increased from 0 V to -10 V, the $\rho_{AHE}$ of the near square-shaped loop was increased by a factor approaching three. When the $V_g$ was below -10 V, the $\rho_{AHE}$ dropped sharply and at $V_g$ = -14 V, the hysteresis loop became absent. The gate-voltage dependent charge density and $R_0/R_S$ ratio are shown in Fig. 2(c). The hole density decreased by $3.4 \times 10^{21}$ cm$^{-3}$ (see Fig. S13 for the definition of carrier density) with the increasing proton intercalation resulting from the gate voltage being swept from 0 V to -14 V. Hence, significant electron doping is induced by the protonic intercalation. The $R_0/R_S$ ratio remained near 1 under different gate voltages. To clarify the evolution of all the AHE loops, the charge density ($n_s$) dependent $\rho_{AHE}$ and coercivity are plotted in Fig. 2(d). With increasing proton intercalation, the $\rho_{AHE}$ at first increases slightly, is maintained until $n_s=2.28\times10^{21}$ cm$^{-3}$ ($V_g$= -6 V), then increases again sharply before once again stabilising from $2.16\times10^{21}$ cm$^{-3}$ to $0.98\times10^{21}$ cm$^{-3}$ ($V_g$ from -8 V to -10 V). Finally, $\rho_{AHE}$ drops from 13.6 μΩ·cm to near zero when the voltage increases to -13 V ($2.8 \times 10^{20}$ cm$^{-3}$). The coercivity declines sharply from 709 Oe to 436 Oe as the hole density decreases from $3.4\times10^{21}$ to $3.34\times10^{21}$ cm$^{-3}$ before rising to a sharp peak of 641 Oe at $n_s=2.28\times10^{21}$ cm$^{-3}$ and undergoing a slight increase from $2.16\times10^{21}$ to $0.72\times10^{21}$ cm$^{-3}$. The coercivity finally vanishes when $V_g$ is swept to -14 V. The complex evolution of coercivity could indicate that the proton intercalation induced metastable magnetic domain structures within the $Cr_{1.2}Te_2$ nanoflake. Note that the coercivity of a vdW magnetic nanoflake is not solely determined by its thickness, but largely by its evolution of magnetic microstructure, which could be further affected by the defects and impurities generated during the mechanical exfoliation, the size of the nanoflake, the competition between the magnetic anisotropic energy and thermal agitation energy, surface stress, crystal dislocation, intralayer and interlayer domain wall motions, etc. Hence, it is common to observe several hundred Oersted coercivity bias between nanoflakes of similar thickness. More gate-voltage dependent AHE loops of device #8 and other devices are shown in Supplemental Material. The absence of hysteresis at $V_g$ = -14 V may be attributed to either a MPT or a change in the magnetic anisotropy. If due to the change in magnetic anisotropy from perpendicular to in-plane, a curved hysteresis characteristic would be observed under a large enough applied magnetic field. As shown in Fig. 2(e), with an applied perpendicular field up to 9 Tesla, the AHE curves at $V_g$ = -14 V of device #8 are nearly flat (Supplemental Materials show similarly near-flat Hall curves under 9 T from other devices). This observation effectively rules out the possibility of altered magnetic anisotropy and substantiates the attribution of a protonic gate induced MPT in $Cr_{1.2}Te_2$.

Since the $Cr_{1.2}Te_2$ nanoflakes exhibit a high Curie temperature (up to 320 K) and a nearly square AHE loops up to T = 290 K, the gate voltage-controlled magnetism of $Cr_{1.2}Te_2$ nanoflakes was investigated further at 290 K. Figure 3(a) shows the protonic gate-voltage dependent AHE curves of device #4 at 290 K. Resembling the results of device #8 in Fig. 2, the $\rho_{AHE}$ firstly increased and reached up to 7.2 μΩ·cm at $V_g$ = -6.5 V, a value 4.5 times higher than that of the pristine material (1.6 μΩ·cm). It then sharply declined, and the hysteresis loop vanished at $V_g$ = -10 V. The figure inset shows the gate voltage dependent hole density, which decreases with increasing proton intercalation, as in device #8. Figure 3(b) shows the gate-voltage dependent coercivity and $R_0/R_s$ ratio at 290 K. As in device #8, the coercivity initially falls to 36.5 Oe, (27% of the coercivity at $V_g$= 0 V), then reaches a sub-peak at $V_g$= -6.5 V before decreasing to 42 Oe at $V_g$= -8.5 V. The evolution of the $R_0/R_S$ ratio shows a similar trend to the evolution of coercivity which drops to around 0.1 at $V_g$= -3 V and -4.5 V and remains around 1 under other gate voltages. Additional AHE data recorded at room temperature from another device (device #5) is shown in the Supplemental Material. Based on the aforementioned experimental results, we conclude that electrically tuned MPTs can be realized up to room temperature in proton-gated $Cr_{1.2}Te_2$ devices.

The complex evolution of AHE in $Cr_{1.2}Te_2$ should result from the evolution of magnetic properties and AHC with proton intercalation. To further explore this aspect of the study, theoretical analysis was carried out using density functional theory (DFT). The computational details are included in Supplemental Material [32] (see also references [33-37] therein). Recently, contradictory reports stated that monolayer $CrTe_2$ exhibited a ferromagnetic ground state [24,38], or an antiferromagnetic ground state [39], probably due to different experimental settings and conditions. Our DFT calculations show that the ground state of monolayer $CrTe_2$ is an intralayer antiferromagnetic state in a zigzag pattern, which can be modelled in a rectangular supercell [see Fig. S20(a)]. As the charge density changed by more than $10^{21}$ cm$^{-3}$ and cross-sectional tunnelling electron microscopy results revealed no significant structural changes with proton intercalation in $Cr_{1.2}Te_2$ (see Fig. S15), the charge density variation is considered a potential factor in causing the MPT. Here, three magnetic configurations were considered, i.e. FM, AFM and ferrimagnetic (FerriM), as shown in Fig. 4(a). The charge doping dependent energy for the three magnetic states are illustrated in Fig. 4(b). From the results, $Cr_{1.2}Te_2$ presents FM state when its hole doping exceeds $1\times10^{22}$ cm$^{-3}$ and the ground state becomes AFM when it is electron doped. Importantly, when the hole doping decreases within the range $n_s$= $10^{22}$ to 0 cm$^{-3}$, the magnetic ground state changes from FM to AFM. This is qualitatively consistent with the experiments. Meanwhile, introducing protons into $Cr_{1.2}Te_2$, in terms of charge, equivalent to injecting electrons (H$^-$) into the system [17], pushes the Fermi level significantly higher (see Fig. S17). Increases of around 0.19 and 1.78 eV result for $Cr_{1.2}Te_2H_{0.2}$ and $Cr_{1.2}Te_2H_2$, respectively, indicating remarkable electronic doping due to the proton intercalation. These results suggest that the proton intercalation-induced electron doping could bias energies of different ground states and thereby contribute to the evolution in the gate-dependent magnetic hysteresis loops. Ultimately, a FM to AFM MPT is induced, as observed in the experiment.

To understand the remanence variation with applied gate voltages shown in Fig. 2 and Fig. 3, the AHC under gate voltages were analysed. A tight-binding Hamiltonian for $Cr_{1.2}Te_2$ was built via the Wannier90 package based on the Cr $d$ and Te $p$ orbitals [40-42]. The AHC was calculated based on the method of Berry curvature, employing the WannierTools codes [43,44]. The calculated band structure, partial density of states (PDOS) and anomalous Hall conductivity of $Cr_{1.2}Te_2$ are presented in Figs. 4(c) 4(d) and 4(e), respectively. The density of states near the Fermi level is mainly determined by the $d$ orbitals of the Cr atoms and the $p$ orbitals of the Te atoms. Significant $\sigma_{xy}$ and $\sigma_{yz}$ AHC components coexist in the material. This phenomenon results from the in-plane magnetization of the doped Cr atoms and the out-of-plane magnetization of the other Cr atoms. This characteristic differs from that observed in ordinary AHC materials, in which the magnetic moments align almost co-linearly along the magnetic easy axis and only one significant AHC component exists [45]. It is obvious that the calculated AHC varies significantly near the Fermi level. As mentioned earlier, the Fermi level is significantly elevated due to proton intercalation. Consequently, the AHC values may be altered. Combining the calculations of AHC in Fig. 4(d) and possible magnetic states in Fig 4(b), we can conclude that the $\rho_{AHE}$ evolution with various gate voltages does not originate from the variation of magnetization but from the large variation of AHC due to appreciable Fermi level shift.

In conclusion, based on experimental results and DFT calculations, we have confirmed an electrically manipulated FM to AFM transition in $Cr_{1.2}Te_2$, a vdW itinerant ferromagnet with a near square-shaped magnetic loop that is maintained up to room temperature. This was achieved using protonic gating, demonstrating the effectiveness of this tool for modulating the magnetic properties of low-dimensional materials. The realization of room temperature electrically-tuned magnetism in vdW magnets with a near square-shaped magnetic loop is a vital step towards the commercial application of vdW magnetic heterostructures-based spintronic devices. With the emergence of AFM spintronics, the observation of an electrically tuned FM-AFM phase transition at room temperature is significant and suggests $Cr_{1.2}Te_2$ nanoflakes as a potential platform for novel AFM spintronic devices.

# Acknowledgements

This research was performed in part at the RMIT Micro Nano Research Facility (MNRF) in the Victorian Node of the Australian National Fabrication Facility (ANFF) and the RMIT Microscopy and Microanalysis Facility (RMMF). L.W. was supported by the Australian Research Council Centre of Excellence in Future Low-Energy Electronics Technologies (Project No. CE170100039), the National Natural Science Foundation of China (Grant No. 12374177, 52072102 and 12235015) and the Funding for Infrastructure and Facility at Hefei University of Technology. Work at South China University of Technology was supported by the National Natural Science Foundation of China (Grant No. 12074126) and the Key Research and Development Project of Guangdong Province (Grant No. 2020B0303300001). Work at High Magnetic Field Laboratory was supported by National Key R&D Program of the MOST of China (Grant No. 2022YFA1602603), the National Natural Science Foundation of China (Grants No. 12274413, U19A2093) and Collaborative Innovation Program of Hefei Science Center, CAS (Grant No. 2022HSC-CIP017). Work at University of Science and Technology of China was supported by Innovation Program for Quantum Science and Technology (2021ZD0302800).

# Figure Legends

FIG. 1. Crystal structure and initial characterization of device #1. (a) Top view and side view of the atomic structure of $Cr_{1.2}Te_2$. The grey dashed line is the bisector of ab axes. Here, Cr-1 is the intralayer Cr atom and Cr-2 represents the interlayer fractional-intercalated Cr atoms. (b, c) Optical and atomic force microscope images of a $Cr_{1.2}Te_2$ nanoflake (device #1) device on a $SiO_2$/Si substrate. The red scale bars represent 5 μm. (d,e) AHE hysteresis loops from 2 K to 200 K. The scales of $\rho_{xy}$ at 2 K and 20 K are zoomed in by 2. (f) AHE curves from 250 K to 320 K, the hysteresis disappeared at 300 K. (g) Temperature dependent coercivity and $R_0/R_s$ values.

FIG. 2. Protonic gating modulation of a 43 nm $Cr_{1.2}Te_2$ nanoflake (device #8) at 200 K. (a) The schematic diagram of a $Cr_{1.2}Te_2$ solid protonic gate device, where a $Cr_{1.2}Te_2$ flake lies on the solid proton conductor and a Pt electrode is utilized as the back-gate electrode. (b) AHE hysteresis loops under various gate voltages. Here only representative loops are plotted to show the major evolution. (c) Gate voltage dependent charge density and remanence. (d) Evolution of $\rho_{AHE}$ and coercivity values with various charge densities. (e) The Hall effect curve of device #8 when $V_g = -14\ V, T = 200\ K$ with a magnetic field scan up to 9 T.

FIG. 3. Voltage-controlled magnetism at 290K in device #4 (43 nm thick). (a) Room temperature AHE loops under different gate voltages. Inset shows the gate voltage dependent charge density. (b) Gate voltage dependent coercivity and $R_0/R_s$ ratio.

FIG. 4. DFT Calculations. (a) Three potential magnetic configurations for $Cr_{1.2}Te_2$. The blue and red arrows represent spin up and down, respectively. (b) Total energy as a function of charge density with three magnetic configurations. Inset is an enlargement of the purple dashed rectangular section corresponding to the experimental range. (c) Band structure of $Cr_{1.2}Te_2$ (d) anomalous Hall conductivity, and (e) PDOS of $Cr_{1.2}Te_2$, respectively.

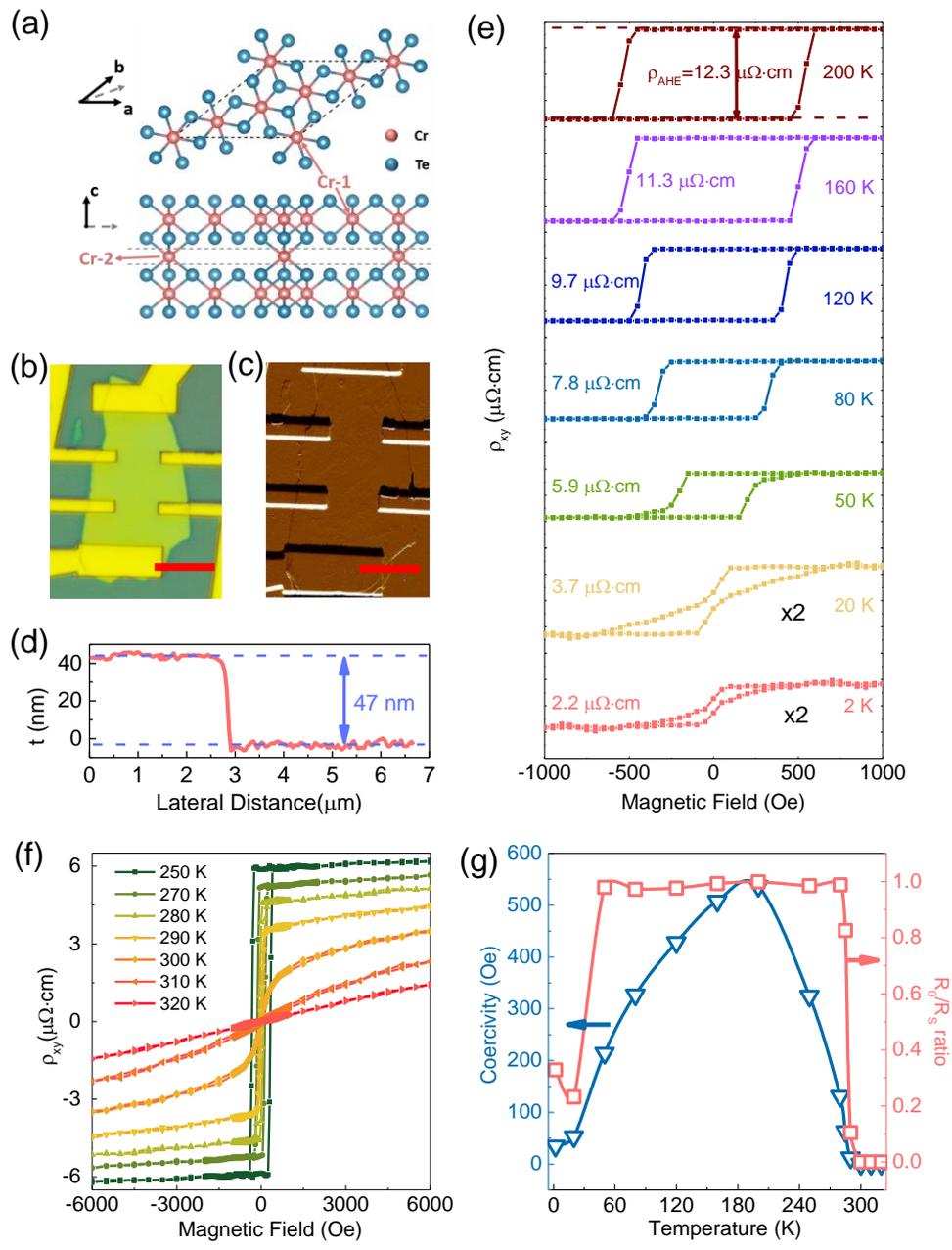

Figure 1

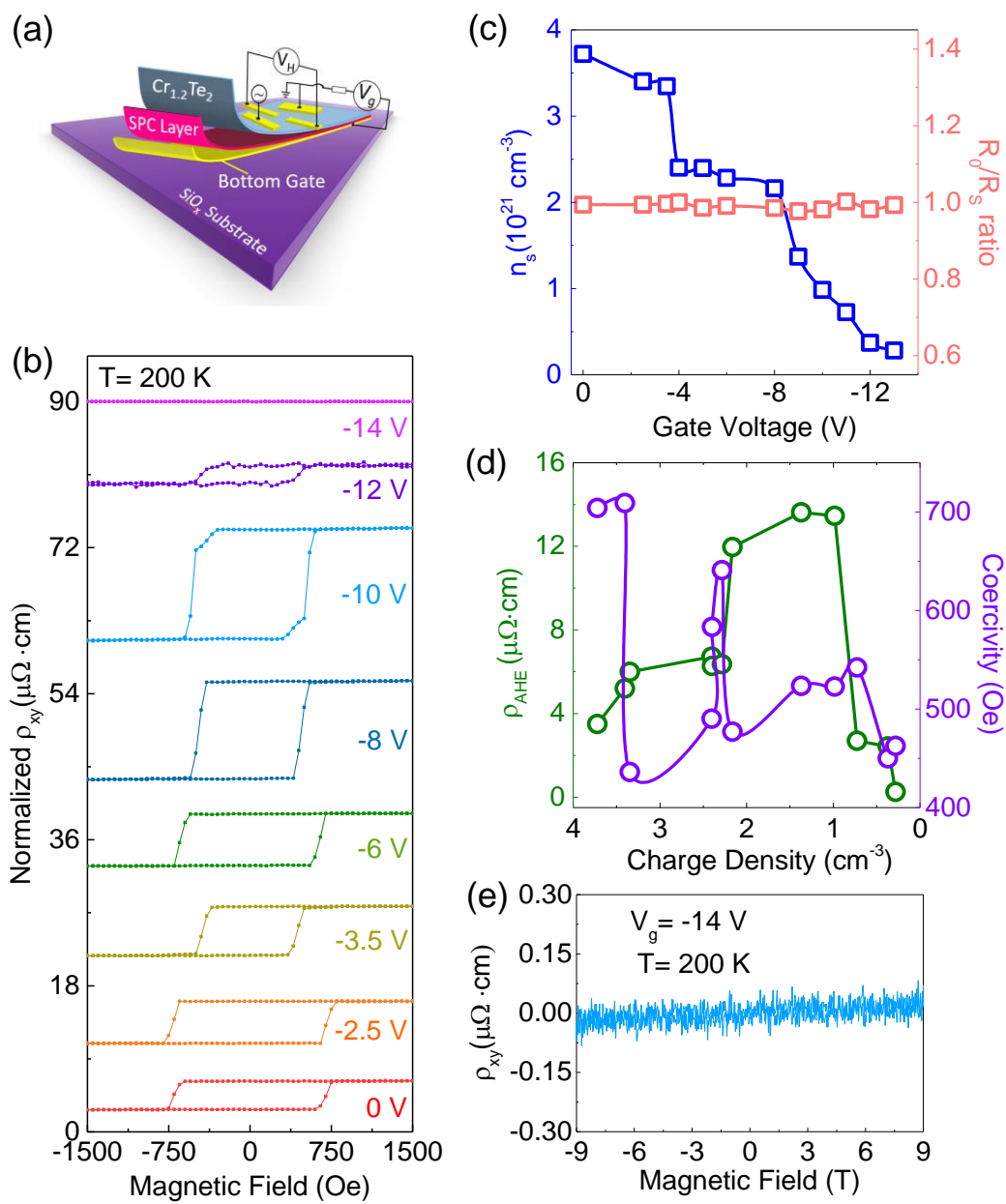

Figure 2

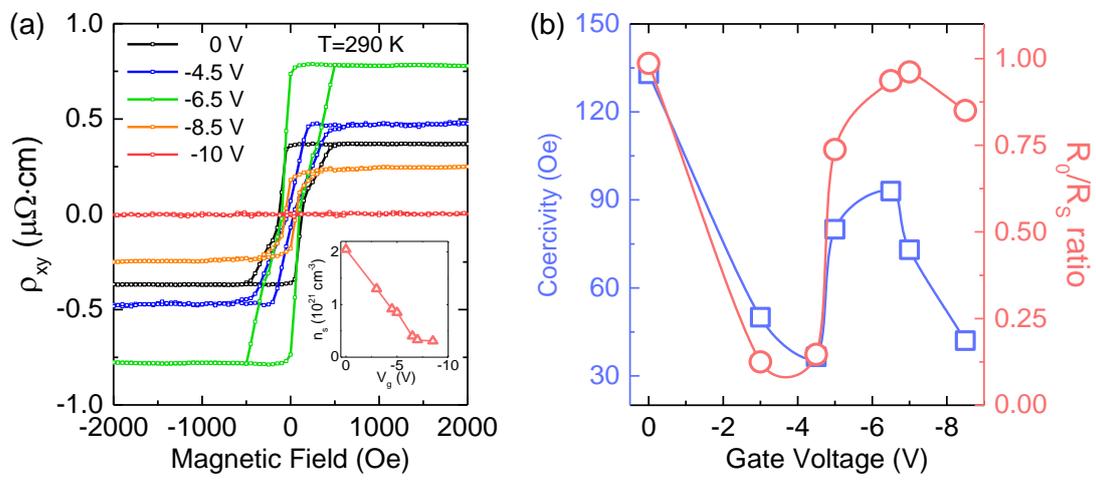

Figure 3

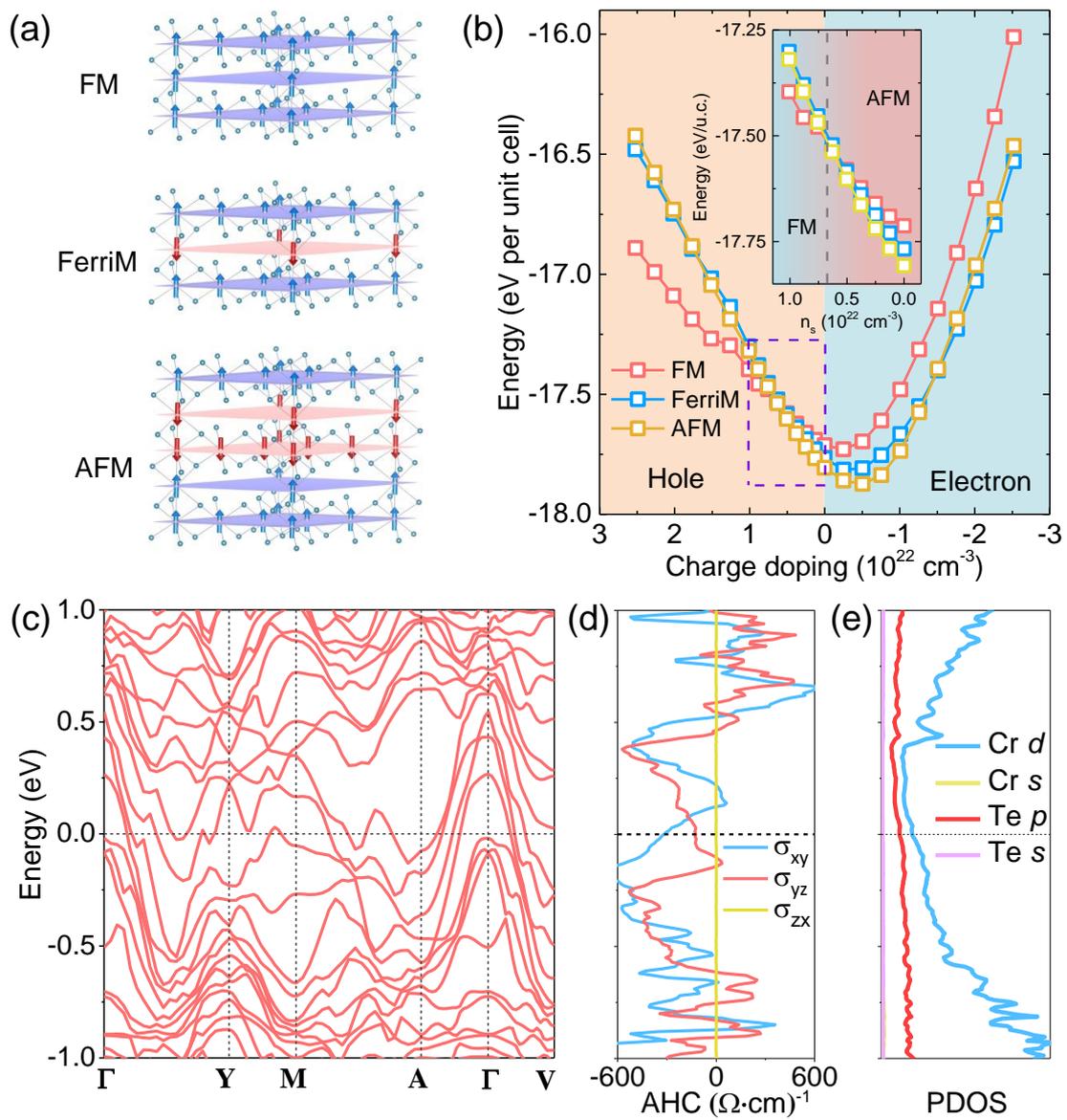

Figure 4